\documentclass[pra,twocolumn,showpacs,amsmath,amssymb,superscriptaddress,longbibliography,reprint]{revtex4-1}
\usepackage{graphicx,float,bm,color,mathptmx,amsfonts,dsfont} 
\usepackage[colorlinks=true,linkcolor=blue,citecolor=blue,urlcolor =blue]{hyperref}

\begin{document}

\title{Simple {factorization} of unitary transformations}

\author{Hubert~de Guise} 
\affiliation{Department of Physics, Lakehead
  University, Thunder Bay, ON P7B 5E1, Canada}

\author{Olivia~Di Matteo} 
\affiliation{Department of Physics and
  Astronomy, University of Waterloo, Waterloo, 
Ontario, Canada N2L  3G1} 
\affiliation{Institute for Quantum Computing, 
University of  Waterloo, Waterloo, 
Ontario, Canada, N2L 3G1}

\author{Luis~L.~S\'{a}nchez-Soto} 
\affiliation{Departamento de \'Optica, Facultad de F\'{\i}sica, 
Universidad Complutense, 28040~Madrid, Spain} 
\affiliation {Max-Planck-Institut f\"ur die  Physik des Lichts, 
Staudtstra\ss e 2, 91058 Erlangen, Germany}

\begin{abstract}
  We demonstrate a method for general linear optical networks that
  allows one to factorize any SU($n$) matrix in terms of two SU($n-1)$
  blocks coupled by an SU(2) entangling beam splitter.  The process
  can be recursively continued in {a straightforward} way, ending
  in a tidy arrangement of SU(2) transformations. The method hinges
  only on a linear relationship between input and output states, and
  can thus be applied to a variety of scenarios, such as microwaves,
  acoustics, and quantum fields.
\end{abstract}

\maketitle

\section{Introduction}

Linear optics constitutes an outstanding setting for information
processing.  The Knill-Laflamme-Milburn~\cite{Knill:2001aa} protocol
for scalable quantum computing, experimental boson
sampling~\cite{Spring:2013aa,Broome:2013aa,Crespi:2013aa,
  Tillmann:2013aa}, or the generation of quantum random
walks~\cite{Do:2005aa,Perets:2008aa,Peruzzo:2010aa,
  Broome:2010aa,Schreiber:2012aa,Bian:2017aa}, are good examples of
how the growing capabilities of fabrication technologies are
transforming the field of quantum photonics~\cite{Carolan:2015aa}.  In
addition, these capabilities are altering classical areas, such as
microwave photonics~\cite{Capmany:2016aa} or optical
networking~\cite{Chen:2011aa,Stabile:2016aa}.

A basic ingredient for all these developments is the design of
reconfigurable setups that can perform any linear operation. The
influential work by Reck \emph{et al.}~\cite{Reck:1994aa}, which can
traced back to the elegant results of
Murnaghan~\cite{Murnaghan:1962aa}, established that a specific array
of basic two-mode operations is sufficient to implement any unitary in
U($n$).  In this way, it is indeed possible to construct a single
device with ample versatility to implement any possible unitary
operation up to the specified number of modes.  Recently,
demonstrations of large-scale linear networks have
appeared~\cite{Harris:2016aa,Harris:2017aa}.

Continued interest in these universal processors for classical and
quantum applications has led to new
designs~\cite{Miller:2015aa,Dhand:2015aa}.  In particular, an
intriguing proposal came out~\cite{Clements:2016aa} requiring roughly
half the optical depth of the original Reck \emph{et al.}
design~\cite{Reck:1994aa}.  This is important for minimizing optical
losses and reducing fabrication resources.

We discuss here a decomposition of any $n\times n$ unitary in terms of
two $(n-1)\times (n-1)$ unitaries coupling the same $n-1$ modes, and a
single $2\times 2$ unitary coupling one of those $n-1$ to the
remaining mode \cite{Murnaghan:1952aa}.  The scheme is recursive; it
can be halted at any dimensionality of subtransformations or performed
in its end, resulting in a tidy arrangement of SU(2) gadgets.  The
structure is thus
\begin{equation}
  R^n(\Omega) =
  R^{n-1}(\tilde{\Omega} ) \,  R_{12}( \alpha, \beta, \alpha) \, 
  R^{n-1} (\tilde{\Omega}^{\prime}) \, .
  \label{factoredform}
\end{equation}
This factorization is economical from a computational perspective: it
requires the evaluation of fewer matrices than that of Reck \emph{et
  al.}~\cite{Reck:1994aa} and this advantage increases with~$n$.  This
economy is particularly relevant as multiparticle scattering by large
unitary arrays are now within the realm of experimental feasibility.
Finally, with the transformations $R^{n-1}(\tilde\Omega)$ and
$R^{n-1}(\tilde\Omega')$ in the same subgroup, the scheme is well
adapted to calculations using the Gelfan'd-Tseitlin
machinery~\cite{Gelfand:1950ihs,Gelfand:1963aa,Alex:2011aa,
Louck:1970aa}.  
  
We demonstrate the universality of the design and explain in detail
some pertinent examples that reveal the directness of the procedure.

As a byproduct, the Haar measure of U($n$) can easily be factorized
according to our scheme. There is a fresh interest in realizing Haar
random unitary matrices~\cite{Zyczkowski:1994aa}, because of the
important role they play in various tasks for quantum
cryptography~\cite{Hayden:2004aa} and quantum
protocols~\cite{Abeyesinghe:2009aa}. From this viewpoint, our
analysis, which is reminiscent of the ideas sketched in
Ref.~\cite{Murnaghan:1952aa}, might be instrumental for a simpler
implementation of these
operations~\cite{Spengler:2012aa,Russell:2017aa}.

Finally, it is important to note that, while our scheme is generally
versatile, applies to any $n$, and can be used for arbitrary
representations of SU($n$), there exist other algorithms in dimension
$2^m$ (see, e.g., Refs.~\cite{Shende:2006,Mottonen:2004,
  Vartiainen:2004aa}) that achieve more efficient decompositions with
respect to the quantum circuit model. Our decomposition does not
improve on the bounds presented in this other work, but instead offers
a convenient and experiment-driven parametrization that retains the
same scaling with $n$ regardless of the internal tensor-product
structure of the system.

\section{Recursive factorization of unitary transformations}

An ideal, lossless linear optical circuit with $n$ input channels and
$n$ output channels performs an optical transformation which can be
described by an $n \times n$ matrix; i.e., it belongs to the group
U($n$).  We can always factor an overall phase to make the determinant
equal to 1, so we deal with SU($n$)~\cite{Murnaghan:1962aa}, which has
$n^{2}-1$ independent parameters.

Our goal is to explore an intuitive factorization of SU($n$)
transformations, which is especially germane for our purposes here and
has the additional advantage of being highly recursive.  To be more
precise, our method can be symbolically stated in the following way:
any $R^n(\Omega) \in$ SU($n$) can be written {as in
  Eq.~(\ref{factoredform})}, where
$R^{n-1}(\tilde{\Omega}), R^{n-1}(\tilde{\Omega}^{\prime}) \in
SU(n-1)$. Here, $R_{ij}$ is a matrix of the form
\begin{equation}
  R_{ij} = 
  \left (
    \begin{array}{ccccccc}
      1 & 0 & \cdots & \cdots & \cdots & \cdots & 0 \\
      0 & 1 & &  & & & \vdots \\
      \vdots &  & \ddots & & & & \vdots \\
      \vdots & & & \mathcal{R}_{ij} & & & \vdots \\
      \vdots & & & &\ddots & & \vdots \\
      \vdots & & & & & 1 & 0 \\
      0 & \cdots & \cdots & \cdots & \cdots & 0& 1 
    \end{array}
  \right ) \,,
  \label{eq:guay}
\end{equation}
coupling adjacent modes $i$ and $j$ (with $j=i+1$) via an SU(2)
transformation $\mathcal{R}_{ij} (\alpha, \beta, \gamma )$ acting on
them.

We recall that any $\mathcal{R} (\alpha, \beta, \gamma) \in$ SU(2),
parametrized by the Euler angles, can be always written as
\begin{align}
  \label{eq:1}
  &  \mathcal{R} (\alpha, \beta, \gamma)   =   
    \mathcal{R}_{z} (\alpha) \, \mathcal{R}_{y} (\beta) \, 
    \mathcal{R}_{z} (\gamma)   = \left (
    \begin{array}{cc}
      e^{i \alpha/2} & 0 \\
      0 & e^{- i \alpha/2}
    \end{array}
          \right )
          \nonumber \\
  & \times 
    \,
    \left (
    \begin{array}{cc}
      \cos ( \beta/2)  & - \sin (\beta/2) \\
      \sin (\beta/2)  & \cos ( \beta  /2) 
    \end{array}
                        \right )
                        \,
                        \left (
                        \begin{array}{cc}
                          e^{i \gamma/2} & 0 \\
                          0 & e^{- i \gamma/2}
                        \end{array}
                              \right ) \, ,
\end{align}
where we follow the standard notation of
Ref.~\cite{Varshalovich:1988ct}.  This factorization is in turn a
prescription for how to construct the SU(2) device: when the
information is encoded in the polarization, a set of three wave plates
is enough~\cite{Simon:1989aa}; for path encoding, this can be mapped
to a beam splitter of transmittance $\cos^{2} (\beta/2) $ and phase
shift $\gamma$, plus a phase shifter that gives the required extra
phase $\alpha$. The action of $R_{ij}$ can also be devised for more
complex systems, such as ion traps~\cite{Shen:2014aa} and
superconducting circuits~\cite{Peropadre:2016aa}.

\begin{figure}
  \centering
  \includegraphics[width=0.80\columnwidth]{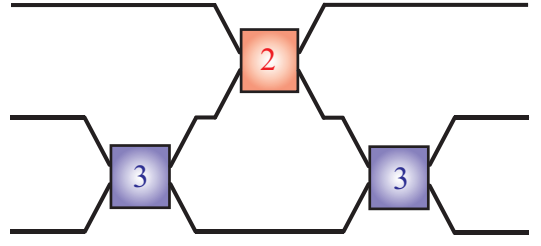}
  \caption{A schematic illustration of the factorization of an SU(3)
    transformation into a sequence of SU(2) transformations. Each mode
    is represented by a line.  Transformations between modes are
    represented by boxes, into which the modes are fed. The number on
    each box indicates the number of parameters in the transformation;
    we use colour for visual ease of distinguishing between
    transformations on the same number of modes, but differing numbers
    of parameters.}
  \label{basicsu3}
\end{figure}

Let us illustrate our scheme in a constructive way, starting with the
simplest case of SU(3). Of course, other parametrizations of SU(3)
elements are possible~\cite{Chang:1985aa,Byrd:1998aa, Ivanov:2006aa},
but one that is particularly useful~\cite{Rowe:1999aa} is into a
sequence of adjacent SU(2)$_{i \, {i+1}}$ transformations mixing
channels $i$ and $i+1$.  More explicitly, with $R^{3}(\Omega) \in$
SU(3), we have
\begin{equation}
  \label{eq:facsu3}
  R^3(\Omega)=R_{23} ( \alpha_1,\beta_1,\gamma_1) \, 
  R_{12}(\alpha_2,\beta_2,\alpha_2) \, 
  R_{23}(\alpha_3,\beta_3,\gamma_3)\, .
\end{equation} 
The middle transformation in the sequence depends only on 2 parameters
(so, it is just a pure beam splitter), and the whole $R^3(\Omega)$
depends on $8$, as it should.  This factorization is symbolically
denoted by a sequence of $2\times 2$ squares representing SU(2)
transformations, as illustrated in Fig.~\ref{basicsu3}.

To lighten the notation, we write $R_{ij}(k)$ where $k$ denotes
the number of parameters in the transformation.  For example,
\begin{align}
  R_{ij}(2) := R_{ij}(\alpha,\beta,\alpha)\, ,
  \qquad
  R_{ij}(3):= R_{ij}(\alpha,\beta,\gamma)
\end{align} 
is used throughout.  In addition, the parameters in the first and last
$R_{23}$ operations are understood to be different even if this is not
indicated in the boxes.  {For completeness we recall that finite
  transformations of the $R_{ij}$ type are obtained by exponentiation
  of generator matrix elements:
  \begin{equation}
    R_{ij}(\alpha,\beta,\gamma)=e^{-i\frac{\alpha}{2} (C_{ii}-C_{jj})}
    e^{-\frac{\beta}{2} (C_{ij}-C_{ji})}e^{-i\frac{\gamma}{2}(C_{ii}-C_{jj})}\, ,
  \end{equation} 
  where $C_{ij}$, with $i,j=1,\ldots n$,  are generators of u($n$) mixing
  modes $(ij)$ when $i\ne j$ or measuring the population $i$ when
  $i=j$.}

%%%%%%%%%%%%%%%%%%%%%%%%%%%%%%%
%%%%%%%%%%%%%%%%%%%%%%%%%%%%%%%%%

To proceed further, we next factorize an SU(4) matrix.  We start with
a $4\times 4$ special unitary matrix $M$ which we write generically as
\begin{equation}
  M=\left(\begin{array}{cccc}
            x&*&*&*\\
            y&*&*&*\\
            z&*&*&*\\
            w&*&*&*\end{array}\right)
\end{equation}
Apply $R^{-1}_{34}(\alpha_1, \beta_1, \gamma_1)$ indicated in
Eq.~(\ref{eq:guay}), namely
\begin{equation} 
{R}^{-1}_{34}(\alpha_1, \beta_1, \gamma_1)=
  \left(\begin{array}{cc}
          \openone_{{2\times 2} }& 0_{2 \times 2 }\\
           & \\
          0_{2 \times 2 } & 
\mathcal{R}^{-1}_{34}(\alpha_1, \beta_1,  \gamma_1)
        \end{array}
      \right) \,.
    \end{equation}
  Choose now the Euler angles as
          \begin{eqnarray}
            & \displaystyle  
              e^{-\frac{1}{2} i (\alpha_1+\gamma_1)} \cos \left(\textstyle\frac{1}{2}\beta_1\right)=
              \frac{z}{\sqrt{1-\vert x\vert^2-\vert y\vert^2}}\, , & \nonumber \\ 
            & \\
            & \displaystyle   
              e^{-\frac{1}{2} i (\alpha_1-\gamma_1)} \sin \left(\textstyle\frac{1}{2}\beta_1\right)=
              \frac{w}{\sqrt{1-\vert x\vert^2-\vert y\vert^2}} \, ,& \nonumber 
          \end{eqnarray}
to obtain
\begin{equation} 
{R}^{-1}_{34} (\alpha_1, \beta_1, \gamma_1)\, M
 = \left(\begin{array}{cccc}
                                   x&*&*&*\\
                                   y&*&*&*\\
                                   \sqrt{1-\vert x\vert^2-\vert y\vert^2}&*&*&*\\
                                   0&*&*&*\end{array}\right) \, ,
\end{equation}
that is, we make a $0$ appear at position $(4,1)$.
%%%%%%%%%%%%%%%%%%
\begin{figure}
  \includegraphics[width=\columnwidth]{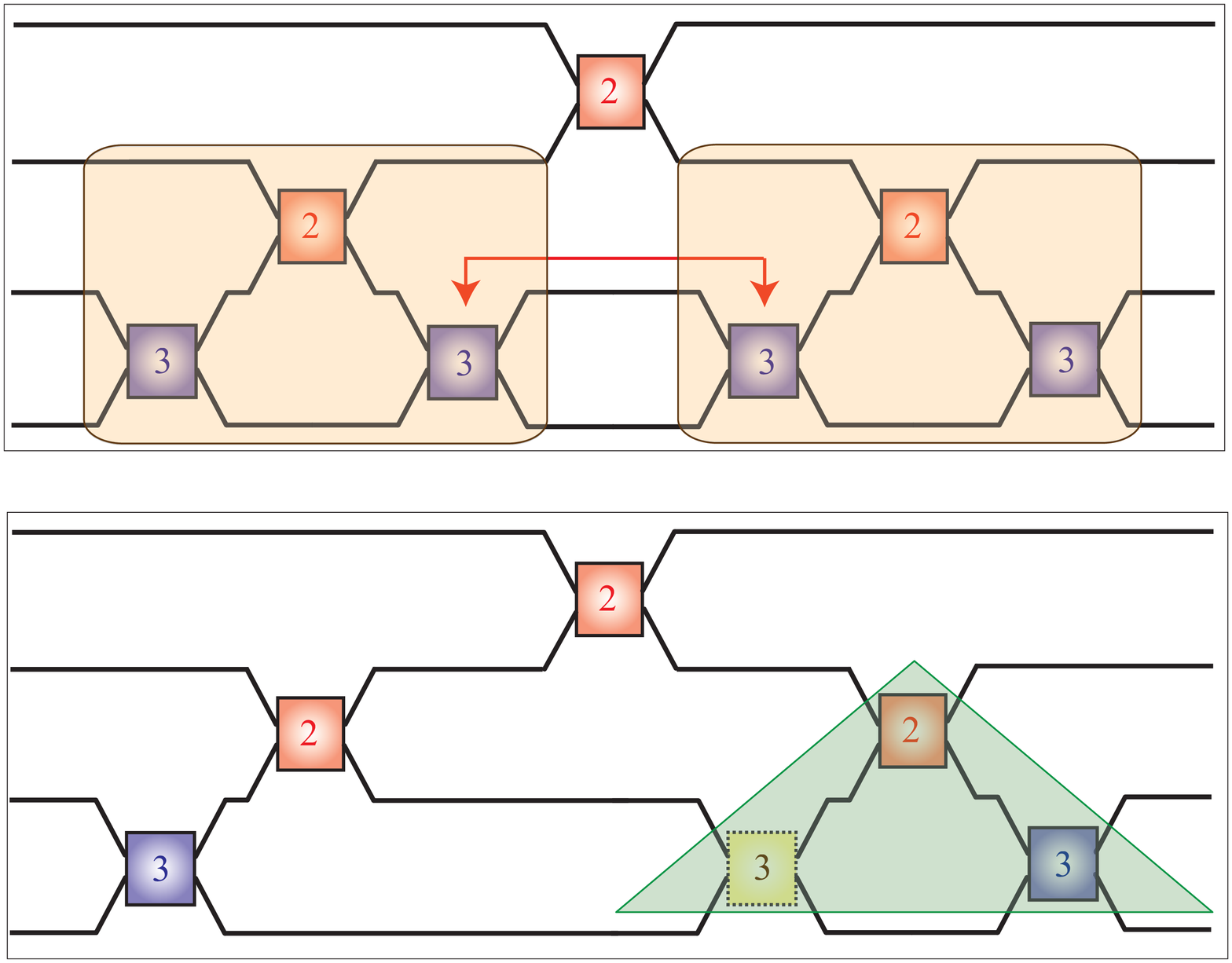}
  \caption{A schematic illustration of the factorization of an SU(4)
    transformation as a sequence of SU(3) transformations, each itself
    written as SU(2) blocks.}
  \label{basicsu4}
\end{figure}
%%%%%%%%%%%%%%%%%%
The second step is apply $R^{-1}_{23}$ to make a $0$ appear at
position $(3,1)$, and finally $R_{12}$ to produce a $0$ in position
$(2,1)$:
\begin{align} 
{R}^{-1}_{12}\, {R}^{-1}_{23} \, {R}^{-1}_{34} \, M &=
   \left(
                                 \begin{array}{cccc}
                                   1&0&0&0\\
                                   0&*&*&*\\
                                   0&*&*&*\\
                                   0&*&*&*
                                 \end{array}
\right)\, ,\\
M&=R_{34}R_{23}R_{12}
   \left(
                                 \begin{array}{cccc}
                                   1&0&0&0\\
                                   0&*&*&*\\
                                   0&*&*&*\\
                                   0&*&*&*
                                 \end{array}
\right)\, , \label{eq:partialsequence}
\end{align}
with the phases chosen so that $1$ occurs in position $(1,1)$.  Since
$\sum_{j=1}^4 \vert a_{ij}\vert^2=1$ for any row of a unitary matrix,
the last step also forces $0$s on the first row.  As all the $R_{ij}$s
are unitary, the result of
$\mathcal{R}^{-1}_{12} \; \mathcal{R}^{-1}_{23}\;
\mathcal{R}^{-1}_{34}$ acting on the original matrix is a $3\times 3$
unitary submatrix, for which the original decomposition in
Eq.~(\ref{eq:facsu3}) can be applied.

Parameter counting {(after suitable relabeling of the modes)} can
be neatly understood graphically.  First, consider an SU(4)
transformation obtained from an SU(2) one of the type
$R_{12}(\alpha,\beta,\alpha)$, sandwiched between two SU(3)
transformations, as illustrated in Fig.~\ref{basicsu4}.  Each SU(3)
transformation is of the type given in Fig.~\ref{basicsu3}, and they
are indicated by shaded squares.

Closer inspection of Fig.~\ref{basicsu4} shows that there are two
adjacent $R_{34}$, joined by a red arrow, that commute with the middle
$R_{12}$, as they mix completely disjoint channels.  One can thus
``push together'' or combine these transformations, as they are of the
\emph{same} SU(2) type, so their combination is a \emph{single} SU(2)
matrix of the $R_{34}$ type. This is symbolically indicated by a box
of different color. The resulting system is in a green shaded
triangle, which represents just a full SU(3) transformation.  The
total number of parameters is 15, as it should be.  Moreover, as a
result of pushing together boxes, the partial SU(3) sequence
$R_{23}R_{34}$ in Eq.~(\ref{eq:partialsequence}) is an SU(3)/SU(2)
transformation obtained from Eq.~(\ref{eq:facsu3}) by setting the
second $R_{23}$ to $\openone$.

\begin{figure}[t]
  \includegraphics[width=\columnwidth]{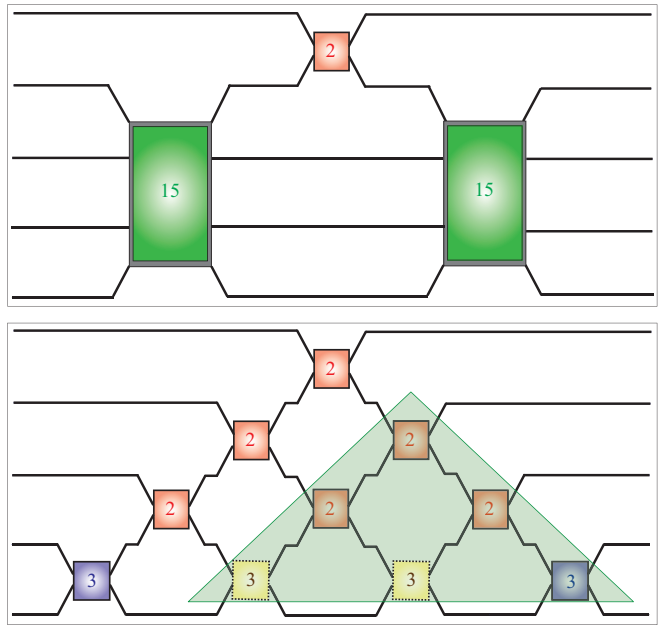}
  \caption{A schematic illustration of the factorization of an SU(5)
    transformation as a sequence of two SU(4) transformations, coupled
    by an SU(2). The bottom panel shows the final result in terms of
    SU(2) blocks. The shaded triangle is an equivalent SU(4)
    transformation.}
  \label{basicsu5}
\end{figure}

We can now immediately generalize the scheme to construct an SU(5)
transformation as an SU(2) sandwiched in between two SU(4)
transformations, represented as green boxes (with 15 parameters) in
Fig.~\ref{basicsu5}.  Again, boxes can be combined into a single SU(3)
following the same pushing procedure.  The final result is just a
SU(4) transformation (indicated again by the shaded triangle) and the
total number of parameters is $24$. We have written a {\sc PYTHON}
software package capable of generating the entire set of parameters,
which we make available online~\cite{Olivia:sun2017}.

%%%%%%%%%%%%
%%%%%%%%%%%%%

At this time it would be useful to compare our decomposition to other
existing schemes.  We recall that factorizations are
representation-independent: even if a scheme is found using the
fundamental $n\times n$ representation of SU$(n)$, it remains valid
for any other representation of SU$(n)$.  Any general SU$(n)$
transformation must also depend on $n^2-1$ parameters: the number of
exponentiations in any scheme must always amount to $n^2-1$ else the
transformation is not general.

\begin{figure}[t]
  \centering
  \includegraphics[width=.95\columnwidth]{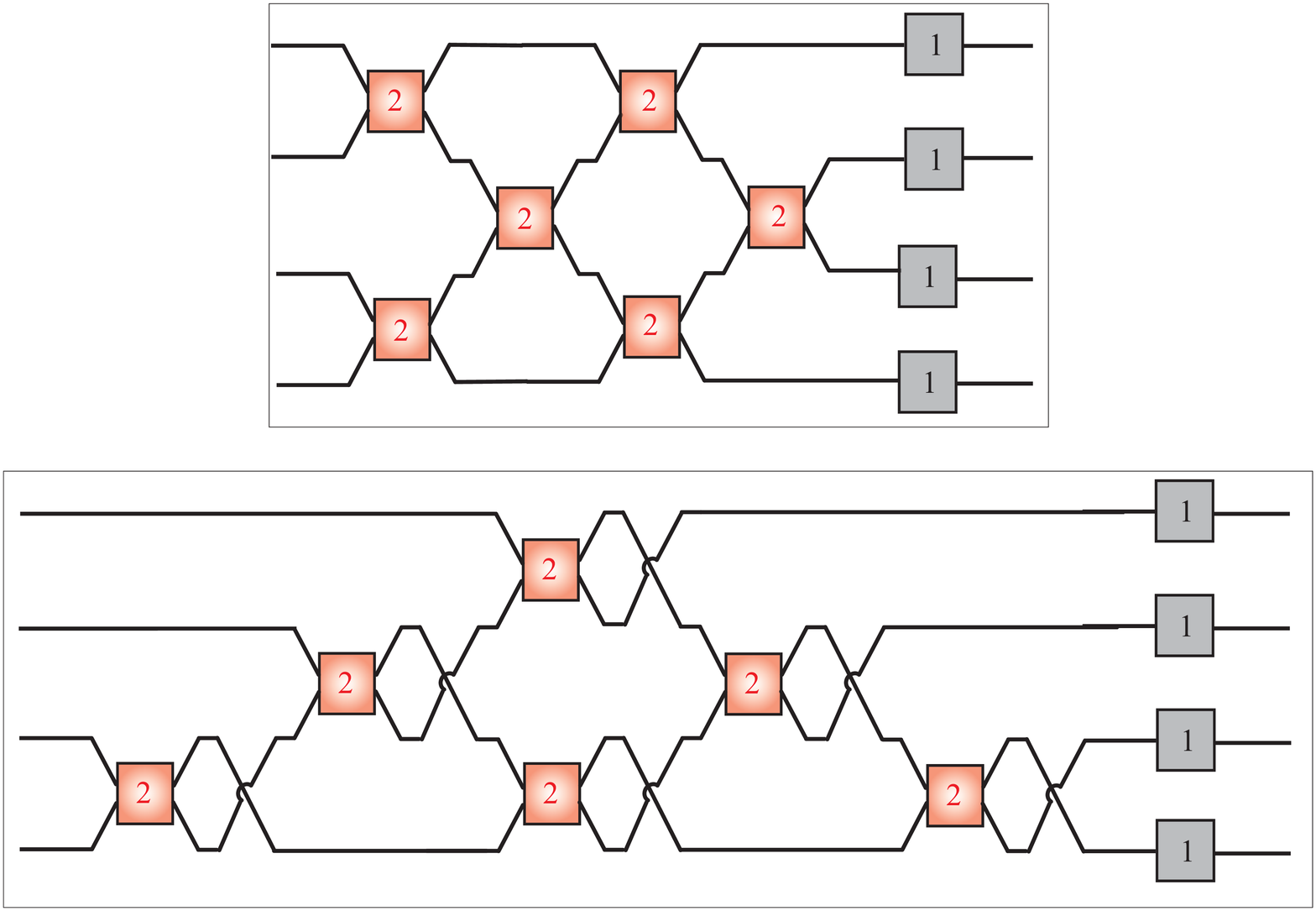}
  \caption{Illustration of the factorization schemes of Clements
    \textit{et al.}~\cite{Clements:2016aa} (top) and Reck \textit{et
      al.}~\cite{Reck:1994aa} for the case of 4 modes.}
  \label{fig:comp}
\end{figure}

In Fig.~\ref{fig:comp} we illustrate the designs of
Reck \textit{et al.}~\cite{Reck:1994aa} and Clements \textit{et
  al.}~\cite{Clements:2016aa} for 4 modes.  Both exclusively employ
2-parameter SU(2) transformations; i.e., the mesh is made only of beam
splitters. The single-mode phase shifts are programmed at the output
of the channels. This is in contradistinction with our results
displayed in Fig.~\ref{basicsu4}.

The decomposition of Reck \textit{et al.}~\cite{Reck:1994aa} uses
transformations on both adjacent and non-adjacent modes, for which the
evaluation of $R_{ij}$ transformations for every possible $(ij)$ pair
of the network is necessary.  The scheme is recursive with SU($n-1$)
transformations easily identifiable as a subblock of the full SU($n$).

Our scheme is also recursive, but with the same type of SU($n-1$)
transformations appearing twice in Eq.~(\ref{factoredform}), and mixes
only adjacent modes.  It achieves computational economy over Reck
\emph{et al.}~\cite{Reck:1994aa} because some generators are used
multiple times, so that fewer of them need to be computed.  For
instance, our SU(4) transformation uses a $(34)$ block three times,
$(23)$ twice and $(12)$ once (in general, $R_{i \, i+1}$ is used $i$
times so $C_{i \, i+1}$ and its transpose conjugate are used $i$
times).

Then our scheme requires 9 types of matrix elements: 6 of the type
$C_{34}, C_{43}, C_{23}, C_{32}, C_{22},C_{33}, C_{44}$ for SU(3)
transformations of modes $(234)$, plus 3 more $C_{12},C_{21}, C_{11}$
for the SU(2) transformation of modes $(12)$. Taking into account the
fact that $C_{ji} = C_{ij}^T$, this generalizes to $n - 1$ matrices of
the type $C_{i \, i+1}$ and $n$ diagonal matrices $C_{ii}$ for
SU($n$). Reck \emph{et al.}, on the other hand, require the evaluation
of three additional matrices for the non-adjacent transformations of
the type $R_{13}, R_{14}$ or $R_{24}$, which entails the computation
of $\frac{1}{2} \left(n^2 - n \right)$ generators of the $C_{ij}$ type
with $j>i$, and $n$ diagonal matrices $C_{ii}$ for SU($n$).  Our
scheme thus saves the evaluation of $\frac{1}{2}(n-1)(n-2)$ generators
over Reck \emph{et al.}~\cite{Reck:1994aa}, with the additional
advantage that the associated scaling in the number of $C_{i \, i+1}$
needed is linear rather than polynomial.  In fact one can see that, by
reusing $(i, i+1)$ blocks, our scheme minimizes the number of matrix
elements to be computed, as one cannot construct a general
transformation by using fewer types of blocks.

%%%%%%%%%%%%%%%%%%%%%%%%%%%%%%%
%%%%%%%%%%%%%%%%%%%%%%%%%%%%%%%%%

This economy becomes very relevant in large networks containing many
particles, as the following pertinent example confirms.  Consider the
scattering of $p$ indistinguishable photons by an $n\times n$
interferometer. This system, currently very popular in the context of
boson sampling, is described by an ${{n+p-1}\choose{p}}$-dimensional
representation of SU($n$) obtained by exponentiating generators using
the same factorization as the fundamental $n\times n$ representation,
with each SU($n-1$) a block diagonal submatrix.  Thus for $n=9$ and
$p=5$~\cite{Wang:2017aa}, one must exponentiate a sequence of matrices
of size $1287\times 1287$.  Permanents are entries of the full
$1287\times 1287$ matrix; i.e., $D$-functions for this irreducible
representation~\cite{Guise:2016aa}.  Whereas the decomposition of Reck
\emph{et al.}~\cite{Reck:1994aa} (or its primal version by
Murnaghan~\cite{Murnaghan:1952aa}) requires the evaluation of $36$
nondiagonal $C_{ij}$ with $j>i$, their transpose conjugates, and $9$
$C_{ii}$, our scheme requires the evaluation of only $8$
$C_{i \, i+1}$ matrices, their transpose conjugates, and $9$ $C_{ii}$.
As the size of practical interferometers increases, the linear scaling
of this scheme thus stands to offer substantial computational savings.
For boson sampling, where the number of modes $n$ is ideally expected
to scale like the square of the number $p$ of photons, the matrices of
the symmetric representation are of size $\sim 10^5\times 10^5$ for
$p=5$.  Clearly, minimizing the number of $C_{ij}$ to evaluate becomes
an issue important from a resource and accuracy perspective.

Note that the factorization of Eq.~(\ref{factoredform}) is also very
natural as the canonical set of basis states, enumerated in terms of
Gelfan'd-Tseitlin patterns $\vert (m)_n\rangle$, also follow the
SU($n$) $\downarrow$ SU($n-1$) subgroup
chain~\cite{Gelfand:1950ihs,Gelfand:1963aa,Alex:2011aa,Louck:1970aa}.
Thus the group functions
\begin{equation}
  \langle (m)_n\vert R^{n-1}(\tilde\Omega)R_{12}(\alpha,\beta,\alpha)R^{n-1}(\tilde\Omega')\vert (m')_n\rangle
\end{equation}
are naturally expressed as a sum of products of SU($n-1$) $\times$
SU(2)$\times$ SU($n-1$) group functions.  A byproduct of this form is
that the SU($n-1$) subgroup transformations are block-diagonal in the
Gelfan'd-Tseitlin basis, a useful feature to check calculations.

The scheme of Clements \textit{et al.}~\cite{Clements:2016aa}
{has a different structure, corresponding instead to a
  rectangular mesh of beamsplitters}. One might expect the triangular
scheme to be more resilient to losses in experiments in which only a
small proportion of its input ports are accessed, whereas the
rectangular scheme is likely to be beneficial for experiments that
involve accessing most of its inputs.

Algorithmically, our scheme differs from the scheme of Clements
\textit{et al.}~\cite{Clements:2016aa} in the order in which 0s are
made to appear when working on the original matrix $M$. As a result
(and by design), the scheme mixes channels ``as early as possible''
and achieves depth of $n$.  In contradistinction our scheme mixes
channels ``as late as possible'': this is necessary to achieve the
highly recursive factorization structure of Eq.~(\ref{factoredform}),
but the tradeoff is a scheme of depth $2n-3$, on par with Reck
\textit{et al.}~\cite{Reck:1994aa}.

This difference in optical depth is the reason why, in a simple loss
model that assumes equal insertion loss for every beam splitter,
Clements \textit{et al.}~\cite{Clements:2016aa} always has better
performance. A careful analysis can be found in
Ref.~\cite{Clements:2016aa}.  In other words, in Clements \textit{et
  al.}~\cite{Clements:2016aa} all the modes encounter roughly the same
number of beamsplitters; in the triangle, transformation $R_{ij}$
occurs $i$ times, then modes experiencing more beam splitters
experience more loss and so the lower modes get more scrambled than
those at the top of the triangle.

Finally, we stress that in our scheme the rightmost $R^{n-1}$
transformation is a full subgroup transformation, while the leftmost
is a partial subgroup transformation.  Pushing and combining boxes
show how an SU($n$) device can be constructed from two SU($n-1$)
devices and a single SU(2) device. In this respect, it is worth
mentioning that the recent interest in networks of multiport devices
instead or beam splitters~\cite{Simon:2016aa,Simon:2017aa} makes our
algorithm especially relevant, as we can decompose a unitary as
coupled SU($d$) devices, with $d$ chosen at will. This makes also the
difference with the well-known decompositions of quantum
gates~\cite{Barenco:1995aa,Cybenko:2001aa,Vartiainen:2004aa}.

\section{Recursive Haar measures}

The recursive factorization in Eq.~(\ref{factoredform}) also implies a
recursive form of the Haar measure. We just briefly recall that a Haar
measure is an invariant measure on the group manifold.  It thus
provides a natural probability distribution over the group, in the
sense that it equally weighs different regions, thus behaving like a
uniform distribution on SU$(n)$.  This is of utmost importance for the
generation of statistical ensembles of unitary
matrices~\cite{Zyczkowski:1994aa}, which is a useful tool in many
fields of physics, as heralded in the Introduction.

For SU(2) we have
\begin{equation}
  d\Omega_2=\sin\beta d\beta d\alpha d\gamma\, .
\end{equation}

Simple application of the usual method yields~\cite{Cornwell:1984aa}
the SU(3) measure, namely,
\begin{align}
  d\Omega_3= d\Omega_2(1) \; 
  \left[\sin\beta_2 \sin^2\left(\textstyle\frac{1}{2}\beta_2 \right) 
  d\alpha_2 d\beta_2\right] 
  \; d\Omega_2(3)
\end{align}
with $d\Omega_2(k)=\sin\beta_kd\beta_kd\alpha_kd\gamma_k$ an SU(2)
measure.

 For SU(4), we find
 \begin{align}
   d\Omega_4&=d\tilde\Omega_3(1,2) \;
              \left[ \sin^4\left(\textstyle\frac{1}{2}\beta_3\right)
              \sin\beta_3\right] \; d\Omega_3(4,5,6) \, ,
 \end{align}
 where $d\Omega_{3} (i, j, k)$ is an SU(3) measure of the arguments in
 parenthesis and
 \begin{equation}
   d\tilde\Omega_3(1,2)
   =d\Omega_2(1)\;
   \left[ \sin\beta_2 \sin^2\left(\textstyle\frac{1}{2}\beta_{2}\right) 
     d\alpha_2 d\beta_2\right] 
 \end{equation}
is a coset measure, with fewer parameters compared to the full
 measure.  The effect of combining $R_{34}$ transformations by pushing
 an $R_{34}$ transformation under $R_{12}$, which we discussed in the
 previous Section, results in the removal of one $d\Omega_2$ factor in
 $d\tilde\Omega_3(1,2)$.

  In SU(5) we find
  \begin{equation}
    d\Omega_5 =d\tilde\Omega_4(1,2,3)
                \;
                \left[ \sin^6\left(\textstyle\frac{1}{2}\beta_4\right) \sin\beta_{4}\right]        d\Omega_4 (5, 6, 7, 8, 9, 10) \, ,
  \end{equation}
  with $d\Omega_{4}$ and $d\tilde\Omega_4(1,2,3)$ having the same
  meaning as before.  The recursion steps to higher $n$ are clear.
  Quite clearly the middle factor is conveniently found to be of the
  form
  \begin{equation}
    \sin\beta_{n-1} \, \sin^{2(n-2)}\left(\textstyle\frac{1}{2}\beta_{n-1}\right)\, ,
  \end{equation}
  with maximum at $\cos\beta_{n-1}=- (n-2)/(n-1)$. This is in
  agreement with the result of \cite{Murnaghan:1952aa} and
  other results obtained from different
  perspectives~\cite{Tilma:2004aa,Russell:2017aa} and it is very
  useful in many instances, e. g., for the parametrization of the
  families of most probable matrices.
  
  The parametrization of Eq.(\ref{factoredform}) and the examples
  above also neatly illustrate how to isolate from the full measure
  the coset measure $d\tilde \Omega_{n-1}$ over SU($n$)/U($n-1$): it
  is obtained by removing the full SU($n-1$) part containing
  $(n-1)^2-1$ factors from full measure.  The usefulness of this coset
  measure comes from applications to coherent
  states~\cite{Zhang:1990aa}; these states ``live'' in the coset space
  SU($n$)/U($n-1$) so the coset measure is what is required for
  integration over these states.

\section{Concluding remarks}

In conclusion, we have discussed the design for universal linear
$n\times n$ optical networks which arises very naturally by recycling
as much as possible the elements already present in a network of size
$(n-1)\times (n-1)$.  Our algorithm decomposes unitary matrices into a
sequence of unitary matrices of one dimension less, entangled by a
beam splitter.  We expect that our compact method will play an
important role in the development of optical processors for both
classical and quantum applications.

In a more technical context, our method allows one to write SU($n$)
group functions in terms of SU($n-1$) group functions, thereby
extending the result of Ref.~\cite{Rowe:1999aa} and the
parametrization of coherent states in SU($n$)/U($m$) for arbitrary
representations of SU($n$) when the highest weight state is
U($m$)-invariant. Work along these lines is now in progress.

\section{Acknowledgments}

The work of H.d.G. and O. D. M. is supported by NSERC of Canada.  IQC
is supported in part by the Government of Canada and the Province of
Ontario. O.D.M. is also grateful for hospitality at the MPL.
L.L.S.S. acknowledges financial support from the Spanish MINECO (Grant
No. FIS2015-67963-P). We thanks L. Rudnicki for helpful discussions.

%\bibliography{elegant}
%\input{AH11412_v2.bbl}


\end{document}